\documentclass[useAMS,usenatbib]{mn2e}
\usepackage{graphics}
\usepackage{epsfig}
\usepackage{amsmath}
\usepackage{color}

\def\lta{\mathrel{\spose{\lower 3pt\hbox{$\mathchar"218$}} \raise
2.0pt\hbox{$\mathchar"13C$}}} \def\gta{\mathrel{\spose{\lower
3pt\hbox{$\mathchar"218$}} \raise 2.0pt\hbox{$\mathchar"13E$}}}
  
\title[X-ray Irradiated Accretion Disk]{The Effect of X-ray Irradiation on the Time Dependent Behaviour of Accretion Disks with Stochastic Perturbations.}
\author[Bari, Misra, Iqbal and  Naveel]{Bari Maqbool$^{1}$\thanks{E-mail:
barispn1@gmail.com}, Ranjeev Misra$^{2}$, Naseer Iqbal$^{1,2}$  and Naveel Ahmad$^{1}$ \\
$^{1}$ Department Of Physics, University of Kashmir, Srinagar-190006, India\\ 
$^{2}$ Inter-University Center for Astronomy and
Astrophysics,  Post Bag 4, Ganeshkhind, Pune-411007, India}

\begin{document}

\date{}

\pagerange{\pageref{firstpage}--\pageref{lastpage}} \pubyear{}

\maketitle

\label{firstpage}

\begin{abstract}
The UV emission from X-ray binaries, is more likely to be produced by reprocessing 
of X-rays by the outer regions of an accretion
disk. The structure of the outer disk may be altered due to the presence
of X-ray irradiation and we discuss the physical regimes where this may occur
and point out certain X-ray binaries where this effect may be important.
The long term X-ray variability of these sources is believed to be due
to stochastic fluctuations in the outer disk, which propagate inwards
giving rise to accretion rate variation in the X-ray producing inner
regions. The X-ray variability will  induce structural variations in the outer
disk which in turn may affect the inner accretion rate.  
To understand the qualitative behaviour of the disk in such a scenario, 
we adopt  simplistic assumptions 
that the disk is fully ionised and is not warped. We develop and use a time dependent global 
hydrodynamical code to study the effect of a sinusoidal accretion rate perturbation introduced
at a specific  radius. The response of the disk, especially the inner accretion rate, to
such perturbations at different radii and with different time periods is shown. 
 While we didn't find any oscillatory or limit cycle behaviour, 
our results show that irradiation enhances
the X-ray variability at time-scales corresponding to the viscous time-scales
of the irradiated disk.

\end{abstract}


\section{Introduction}

Accretion disks are formed when matter from a companion star accretes on to
a compact object which is  usually a black hole or a neutron star. The structure
of such disks can be described in the ``$\alpha$-disk" prescription \citep{Sha73} which
is  often referred to as the standard accretion disk model. Most of the energy
is released in the inner regions of the disk and is radiated at X-ray energies
and hence these systems are observed as X-ray binaries. The flux generated and
the temperature of the disk decreases with radius and hence the outer regions of
the disk emit in UV or in the optical. However, there are couple of ways by which
the  outer region of the disk effects the X-ray emission from the inner parts. 
If the outer region is non-ionised the disk becomes unstable. This hydrogen
ionised thermal instability leads to a  long term limit cycle like variability
in the accretion rate \citep{Cann84} . This is believed to be the origin of X-ray
novae where an X-ray binary rises from quiescence on a time-scale of days and subsequently
its luminosity decays in months time-scale. X-ray novae typically recur in time-scales
of decades. For persistent X-ray binaries, the outer disk is perhaps always ionised
and hence such systematic long term variation is not seen. However, even for such persistent
sources the X-ray emission displays stochastic variability in a wide range of time-scales   
ranging from milli-seconds to months. While the short term variability should arise
from the inner regions, it is the outer regions which is thought to be responsible
for the long term variation. A popular model for the long term variability is the
stochastic fluctuation model \citep{Lyu97} where fluctuations in different radii of the
disk, induce accretion rate variations in the viscous time-scale of that radius. These
accretion rate variations then propagate inwards and cause the observed long term X-ray
variability. Thus, the outer region of an X-ray disk effects the X-ray emission from the
inner parts by regulating the accretion rate inflow into the inner region.

On the other hand, as noted and described by \cite{Cunn76}, the X-ray irradiation by the
inner regions can affect the structure of the outer one.  
There have been several works to understand the structure
and the resultant spectra 
of the outer regions in the presence of X-ray irradiation \citep[e.g.][]{Haya81,Hoshi88,
Tuch90,Yuan91, Dub99, Ivan04, Wick05,Ritt08,Mesh11}. The reprocessed emission will
be emitted in UV/optical and the spectra of such disks has been calculated and compared to
observations \citep{Vrt90,Haya81,Ritt08}. The detailed vertical structure of such
disks which depends on the radiative and heat transfer processes have been studied to
better understand their spectral properties \citep{Ivan04,Wick05,Tuch90,Soon99,Mesh11}. 
The global
structure of such irradiated disks is also expected to be complex with several studies
showing that the disk may be warped or inflated  \citep{Haya81,Dub99}. The effect on the
stability of the disk, especially on the thermal hydrogen ionised instability during
an X-ray novae outburst has been studied  \citep{Yuan91,Ritt08} and found to be important.

Thus, this is an interesting situation where, while the accretion rate in the
inner regions (and hence the X-ray flux)  can be modulated by the outer disk, the
outer disk itself can be strongly affected by the variations in the X-ray flux. In such
systems involving feedback, there are possibilities of characteristic behaviour such as
limit cycle oscillations, resonance or enhanced variability. We
study here these effects by computing the response of such an X-ray irradiated disk to
perturbations, using a global time dependent hydrodynamic code.  For simplicity
and to understand the basic phenomenon, we restrict the analysis to fully ionised
disks without warps. 

In the next section, we describe the steady state X-ray irradiated disks and quantify
the regions of such disk where irradiation changes the disk structure. Using the known orbital
properties of some X-ray binaries, we estimate whether their outer disks are expected to
be influenced by X-ray irradiation. We then proceed in \S 3 to describe the time dependent
accretion disk equations and present the results of the computation in \S 4. In \S 5 we
summarise and discuss the work.

\begin{figure}
\begin{center}
{\includegraphics[width=1.0\linewidth,angle=0]{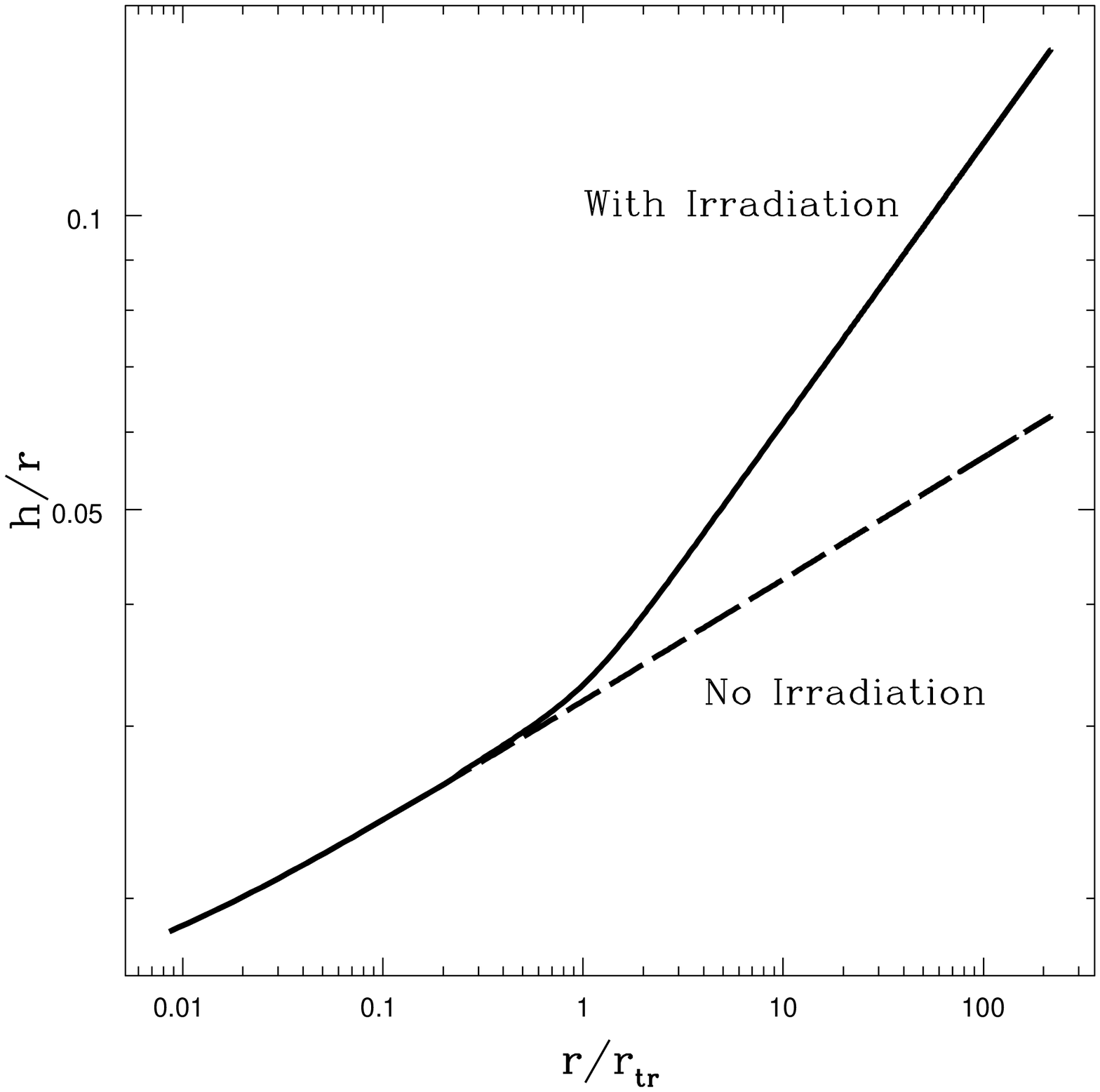}}
{\includegraphics[width=1.0\linewidth,angle=0]{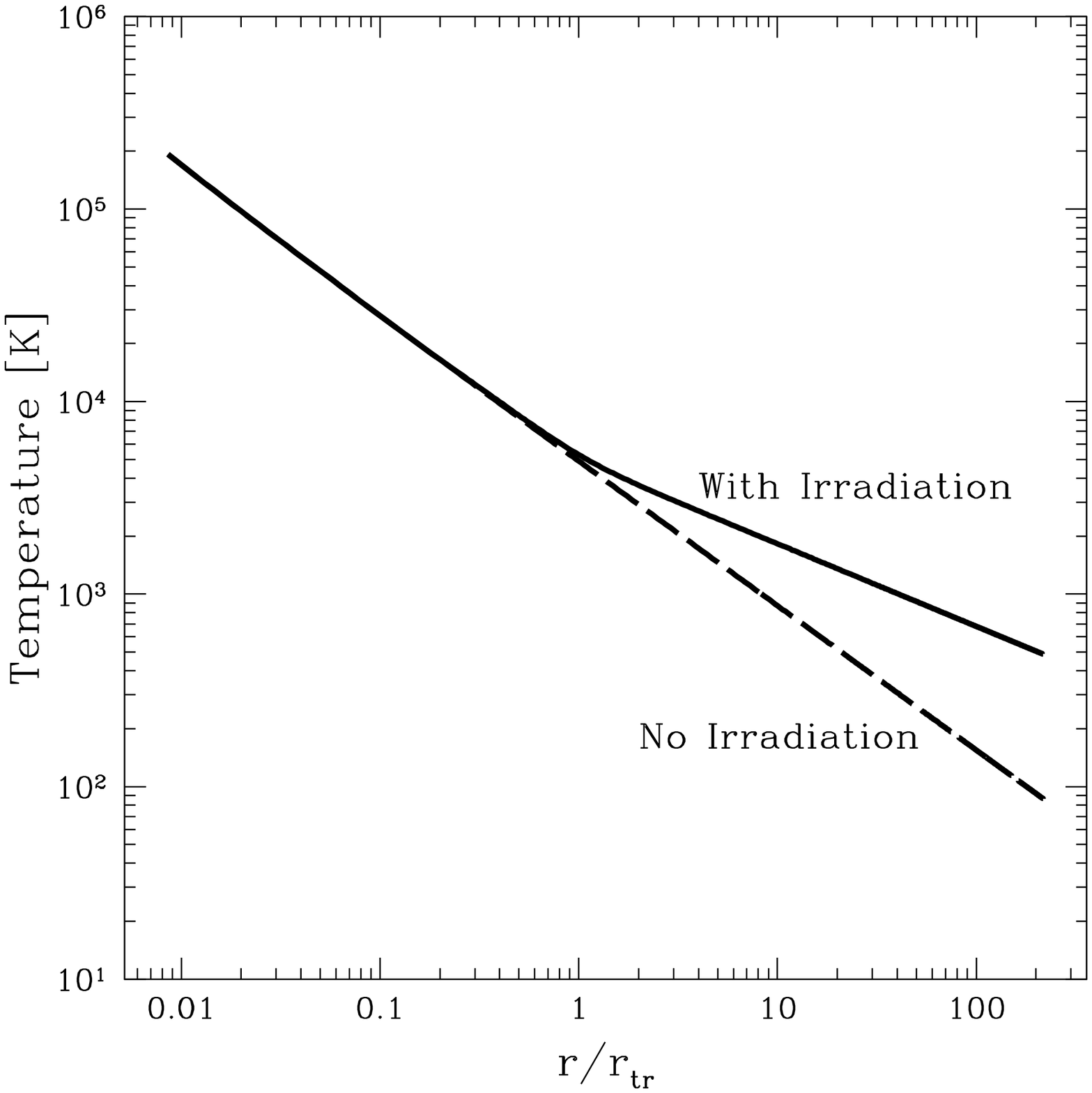}}
\end{center}
\caption{The normalised scale height ($h/r$) of the accretion disk versus  radii normalised to
the transition radius obtained using fixed values of $\dot{M}~=~10^{17}$ gs$^{-1}$, $\alpha~=~0.1$ and $\eta~=~0.05$.
The solid line corresponds to the case when X-ray irradiation is
considered while the dotted line is when the effect is not considered. For radii greater than the
transition radius the disk has a larger height and temperature for the case when irradiation 
is considered as compared to when it is not. Note that the transition to the irradiation dominated
region is fairly sharp.  }
\label{hrb}
\end{figure}

\section{Steady State structure of X-ray Irradiated Accretion Disks}

In the $\alpha$-prescription used in the standard accretion disk model \citep{Sha73}, the
angular momentum transfer equation is given by
\begin{equation}
\frac{\partial{}}{\partial{r}}\left[4 \pi \alpha P r^{2}h\right] ~=~\dot{M} \frac{\partial{}}{\partial{r}}\left[(GMr)^{1/2}\right]
\label{timedep}
\end{equation}
where $M$ is the black hole mass, $\dot M$ is the accretion rate, $r$ is the radius and $\alpha$ is the
viscous parameter. For steady state, when $\dot M$ is a constant and independent of radius, the equation
can be integrated to give, 
\begin{equation}
(\alpha P)(2\pi r \times 2h)(r) = {\dot M}(GMr)^{1/2}
\label{amc}
\end{equation}
In order to study the structure of the outer disk, we neglect
the inner boundary condition and consider the pressure, $P ={2\rho kT}/{m_{p}}$ to
be the ionised pressure of a gas at temperature $T$ and density $\rho$.
The half-thickness of the disk, $h$ can be estimated using hydrostatic equilibrium
in the vertical direction to be
\begin{equation} 
\frac{P}{\rho} ~ \sim ~ \frac{2 k T}{m_{p}} ~ \sim ~ \left(\frac{GM}{r^{3}}\right)h^{2}
\label{heqe}
\end{equation}

The gravitational energy generated per unit area of the disk is given by
\begin{equation}
F_o (r)  =\frac{3\dot{M}}{8\pi r^{2}}\frac{GM}{r}
\end{equation}
and in the standard disk, this is balanced by the radiative energy transfer in the vertical
direction i.e. $\sim \sigma T^4/\tau$, where $\tau$ is the optical depth given by
the surface density $\Sigma = \rho h$ times the opacity $\kappa $. In the presence of
X-ray  irradiation, there is an additional heating term $F_{ir}$. In the special case,
when $F_{ir} >> F_o$, and if one considers the disk to be vertically iso-thermal, the
radiative flux should be $\sigma T^4 \sim F_{ir}$ \citep{Vrt90}. Here we follow 
\cite{Dub99} to write a general equation of the form
\begin{equation}
\sigma T^{4}=F_{0}(r)[\Sigma(\bar{\kappa}_{ff}+\bar{\kappa}_{es})]+F_{ir}(r,h)
\label{mode}
\end{equation}
where we have considered both the opacity due to electron scattering $\bar{\kappa}_{es}$ and
due to free-free interaction $\bar{\kappa}_{ff}$. 
Opacity due to free-free interaction
is given by Kramer's opacity law which is written as
\begin{equation}
\bar{\kappa}_{ff}~=~0.64 \times 10^{23} \rho T^{-7/2}~cm^{2}~g^{-1}
\end{equation}
The heating due to X-ray irradiation is
given by \citep{Vrt90}
\begin{equation}
F_{ir}(r,h)=\frac{f_{2}L_{x}}{4\pi r}\frac{\partial h/r}{\partial r}
\end{equation}
where $f_{2}$ is the absorbed fraction of the radiation impinging on the surface 
and $L_{x}$ is the X-ray luminosity which can be written as $L_x = \eta \dot{M}_{in}c^{2}$,
where $\eta$ is the radiative efficiency of the accretion and $\dot{M}_{in}$ is the accretion 
rate in the inner region. For steady state, $\dot{M}_{in}$ is the same constant accretion rate
at every radii, but we denote it by a different symbol here for latter convenience.

The above equations can now be re-written to provide a differential equation after transforming
$r = xr_g$ and $h = y r_g$  ($r_g = GM/c^2$) to,
\begin{equation}
\frac{\partial{y}}{\partial{x}}=\frac{y}{x}+A_{p}x^{-10}y^{8}-B_{p}x^{25/2}y^{-12}-C_{p}x^{1/2}y^{-2}
\label{h_r}
\end{equation}
Where
\begin{equation}
A_{p}=\frac{4\pi \sigma r_{g}^{2}c^{6}}{f_{2}\eta\dot{M}_{in}}\left(\frac{m_{p}}{2k}\right)^{4}
\label{Ap}
\end{equation}
\begin{equation}
B_{p}= 6 \times 10^{21}\left(\frac{\dot{M}^{3}}{\dot{M}_{in}r_{g}^{3}c^{9}}\right) \left(\frac{1}{\pi^{2}\alpha^{2}f_{2}\eta}\right)\left(\frac{2k}{m_{p}}\right)^{7/2}
\label{Bp}
\end{equation}
\begin{equation}
C_{p}=0.375 \left(\frac{\dot{M}^{2}}{\dot{M}_{in}r_{g}c}\right)\left(\frac{\bar{\kappa}_{es}}{\alpha \pi f_{2}\eta}\right)
\label{Cp}
\end{equation}

This differential equation can be solved to obtain the height as a function of radius and 
subsequently one can obtain the temperature $T$ and the surface density $\Sigma$. When
irradiation dominates the gravitational energy release ($F_{ir} >> F_o$) then the last
two terms can be neglected and the solution is $y \propto r^{9/7}$ corresponding to
the solution reported by \cite{Vrt90}. In the opposite regime when, $F_{ir} << F_o$, then
there are two cases. If the opacity is dominated by scattering, $\kappa_{es} >> \kappa_{ff}$,
then the equation reduces to $A_{p}x^{-10}y^{8} = C_{p}x^{1/2}y^{-2}$ and when 
$\kappa_{es} << \kappa_{ff}$, it becomes $A_{p}x^{-10}y^{8}= B_{p}x^{25/2}y^{-12}$.
\begin{table*} 
\caption{Comparison between the Circularization, Tidal and Transition Radii for some Low Mass X Ray Binaries.}

\begin{tabular} {l c c c c c c c c}
\hline  
Name & Type & $P_{orb}$  & $M_{1}$ &$M_{2}$& $R_{c}$ & 
$R_{T}$ & $r_{tr}$ & $T$ \\ 
& & (d) &$M_{\odot}$ &$M_{\odot} $&($10^{11}cm$) & 
($10^{11}cm$) & ($10^{11}cm$) & (K) \\

\hline

4U 1811-17$^{*}$ & NS & $24.0667$ & $1.4$ &$5$& $3.5$ &  $7.0 $ & $2.2$ & $4900$ \\
3A1516-569$^{*}$ & NS & $16.60$ & $1.4$ &$4$& $2.7$&  $5.4 $ & $2.2 $ & $4900$  \\
CygX-2$^{**}$ & NS &  $9.84 $& $1.780$ &$0.63 $& $2.5 $&   $5.0 $ & $2.9 $ & $4400$ \\
V395 CAR$^{**}$ & NS & $9.02$ & $1.44$ &$0.35 $& $2.7 $ & $5.4 $ & $2.3$ & $4800$ \\
\hline
GS2023 + 338$^{*}$ & BHC & $6.4750$ & 12 &$0.6 $& $4.1 $ & $8.2 $ & $33.7$ & $3800$ \\
GRO J1655-40$^{**}$ & BHC & $12.73$ & $7$ &$2.3$ & $1.07$ & $2.14$ & $15.7$ & $4700$\\
XTE J2123-058$^{**}$ & NS & $0.25$ & $1.415$ &$0.53$& $.22 $ & $0.44$ & $2.2 $ & $4900$ \\
2A 1822-371$^{**}$ & NS & $0.23$ & $0.97$ &$0.33$& $0.21 $ & $0.42 $  & $1.4$& $5700$ \\
GRS 1915 +105$^{*}$ & BHC & $33.500$ & $14$ &$0.81$& $11.7$ & $23.4 $ & $40.6$ & $3500$ \\
2A 1822-371$^{**}$ & NS& $1.7$ & $1.5$ & $2.3$ & $0.6$ & $1.2$ & $2.4$ & $4700$ \\
\hline
\end{tabular}

\begin{tabular}{l}
$M_{1}$ is the mass of the central compact object, $M_{2}$ is the mass of the companion star, $R_{c}$ is the Circularization radius,\\
$R_{T}$ is the Tidal radius and $r_{tr}$ is the transition radius of the accretion disk. The temperature $T$ is the temperature of \\
the accretion disk at the transition radius. NS - Neutron Star. BHC - Black Hole Candidate. \\
\**\citep[] [and references therein]{Zha11},\***\citep {lmxbcat}.
\end{tabular}
\label{table1}
\end{table*}
Solving the differential equation for $\dot{M}~=10^{17}$~g/s,
$\alpha=0.1$, and $\eta=0.05$, one obtains the scale height  and
the mid-plane temperature (Fig. \ref{hrb}). The outer radii where
X-ray irradiation dominates has a higher temperature and scale height
compared to what it would have been if there was not X-ray irradiation.
The transition from viscous dominated to X-ray irradiated is rather sharp
and the radius, $r_{tr}$ where this occurs can be estimated from equating
$F_o(r_{tr}) \Sigma \bar \kappa_{ff} \sim F_{ir} (r_{tr})$ which gives

\begin{equation}
r_{tr}\sim 2\times10^{11}~cm \left(\frac{\dot{M}}{10^{17}g/s}\right)^{2/45}\left(\frac{M}{1.4M_\odot}\right)^{11/9}\left(\frac{\alpha}{0.1}\right)^{-28/45}
\label{rtr}
\end{equation}
where we have assumed (as is typically the case) that $\bar \kappa_{ff} >>
\bar \kappa_{es}$. Note that the transition radius is insensitive to
the accretion rate ( $\propto \dot{M}^{2/45}$). The temperature of the disk at the transition
region is given by

\begin{equation}
T(r_{tr}) \sim 4.9\times 10^{3}~K~\left(\frac{\dot{M}}{10^{17}g/s}\right)^{2/7}\left(\frac{M}{1.4M_\odot}\right)^{-3/7}
\label{temp_tr}
\end{equation}
Since we will be interested in the temporal behaviour of these disks,
the viscous time-scale i.e. the radius divided by the local radial
speed at the transition radius,
\begin{eqnarray}
(T_{visc})_{tr}&=&4\times 10^{7} secs \times \left(\frac{\dot{M}}{10^{17}g/s}\right)^{-11/45} \nonumber \\
&&\left(\frac{\alpha}{0.1}\right)^{71/45} \left(\frac{M}{1.4M_\odot}\right)^{16/9}
\label{tvisc}
\end{eqnarray}
sets a time-scale for any variability.
 
\subsection{X-ray binaries with irradiated disks}

An X-ray binary will have an X-ray irradiation dominated outer disk
provided the transition radius (Eqn \ref{rtr}) is smaller than the
extent of the disk. In a binary, as the matter flowing from the secondary to the
primary possesses angular momentum, it will settle to a radius called the circularization
radius which is  given by \citep{Fra02}
\begin{equation}
R_{circ} = 4(1+q)^{4/3} [0.500 - 0.227\log_{10} q]^{4} P_{d}^{2/3}R_{\odot}
\label{circ}
\end{equation}
where $q$ is the ratio of masses of the companion star and the compact
object and $P_{d}$ is the orbital period in days. Due to viscosity, angular
momentum is transported outwards causing the disk to be larger than
the circularization radius until it is disrupted by the tidal forces
of the companion star. Typically, the Tidal radius $R_{T}$ is related to the
Circularization radius as $R_{T} \sim 2 R_{circ}$  \citep{Shah98}. Thus
only systems with the transition radius $r_{tr} < R_T$, will have an
outer disk whose structure is affected by X-ray irradiation.

We selected ten X-ray binaries from \citet{lmxbcat} and \citet{Zha11}
for which the orbital period and the masses of the primary and companion 
are well constrained and list them in Table \ref{table1}. There are three black hole
candidates (BHC) and seven neutron star (NS) systems in the sample. We computed the
circularization and tidal radii  and compared it with the transition one. We have computed the 
transition radius for $\alpha~=$ 0.1, and accretion rates $10^{17}$ g/s and $10^{18}$ g/s for Neutron
stars and black holes respectively. We find
that for four of these sources (listed in the above panel of the table) the transition
radius is smaller than the tidal one, indicative that these systems may have
an outer disk whose structure is affected by X-ray irradiation. We also computed
the temperature at the transition radius.
Typically the temperature is not high enough to justify that the
disk is completely ionised, which has been discussed in the last section.

\section{Time-Dependent accretion disk with X-ray irradiation}
The surface density $\Sigma$ will change with time if the accretion rate varies with radius and its behaviour
is governed by the conservation of mass equation
\begin{equation}
\frac{\partial{\Sigma}}{\partial{t}} ~=~ \frac{1}{2\pi r_{g}^{2}}\frac{1}{x}\frac{\partial{\dot{M}}}{\partial{x}} 
\label{timedepA}
\end{equation}
The angular momentum transfer equation (\ref{timedep}) provides the accretion rate as a function of
$\Sigma$ and its derivative
\begin{equation}
\dot{M}~=~H\left[-\Sigma x^{-3/2}y^{2} + 2\Sigma \frac{\partial{y}}{\partial{x}} x^{-1/2}y + \frac{\partial{\Sigma}}{\partial{x}} x^{-1/2}y^{2}\right]
\label{mdot}
\end{equation}
where $H = 4 \pi \alpha (GMr_{g})^{1/2}$.
Combined with Equation (\ref{timedepA}), this provided a second order partial
differential equation in $\Sigma$, a diffusive type equation describing the temporal behaviour of the
disk. The normalised scale height $y = h/r_g$ is obtained from  Equation \ref{h_r}, written in terms of
$\Sigma$, i.e.
\begin{equation}
\frac{\partial{y}}{\partial{x}}=\frac{y}{x}+A^{'}_{p}x^{-10}y^{8}-B^{'}_{p}\Sigma^{3}x^{8}y^{-6}-C^{'}_{p}\Sigma^{2}x^{-5/2}y^{2}
\label{h_rp}
\end{equation}
Where
\begin{equation}
A^{'}_{p}=\frac{4\pi \sigma r_{g}^{2}c^{6}}{f_{2}\eta\dot{M}_{in}}\left(\frac{m_{p}}{2k}\right)^{4}
\label{App}
\end{equation}
\begin{equation}
B^{'}_{p}= 6.4 \times 10^{22}\left(\frac{6 \pi \alpha}{f_{2} \eta c^{6} \dot{M}_{in}}\right)\left(\frac{2k}{m_{p}}\right)^{7/2}
\label{Bpp}
\end{equation}
\begin{equation}
C^{'}_{p}=6 \left(\frac{c r_{g}}{\dot{M}_{in}}\right)\left(\frac{\bar{\kappa}_{es} \pi \alpha}{f_{2} \eta}\right)
\label{Cpp}
\end{equation}
In these equations, the local accretion rate  $\dot M (x,t)$ is different from the accretion rate at the inner
radius $x_{in}$, denoted by $\dot M_{in} (t) = \dot M (x_{in},t)$.  

We solve the above set of differential equations using standard techniques and by imposing the courant condition 
$\frac{2D\Delta{t}}{(\Delta{x})^{2}} \leq 0.5$ for stability, where $\Delta t$ is the variable time step,  $\Delta x$ is the bin size
in the radius and  $D$ is the approximate diffusion coefficient of the time varying equation. We tested the code by evolving
for several times the viscous time-scale of the outer radius and found the structure of the disk to be stable after a
short transient phase. Next, we increase the accretion rate at the outer radius to twice its initial value and as
expected after evolving for a time longer than the viscous time, the disk settled into a structure identical to
the steady state for the enhanced accretion rate. The only issue, was that at small radii, when the disk is 
dominated by viscous heating (i.e. X-ray irradiation is not important), then the last three  terms in Eqn \ref{h_rp}
are large, making the derivative $ \frac{\partial{y}}{\partial{x}}$, susceptible to numerical errors. In such cases,
we use instead of Eqn \ref{h_rp}, the expression for the height of the disk when X-ray irradiation is not important
i.e.
\begin{equation}
A^{'}_{p}x^{-10}y^{8}-B^{'}_{p}\Sigma^{3}x^{8}y^{-6}-C^{'}_{p}\Sigma^{2}x^{-5/2}y^{2}~=~0
\label{h_nonir}
\end{equation}
We do this change over when the scale heights computed by Eqn (\ref{h_rp}) and Eqn (\ref{h_nonir}) differ by less than
5\%. This scheme allows us to obtain in general, well behaved solutions both in time and in radius.
 \begin{figure}
\begin{center}
{\includegraphics[width=1.0\linewidth,angle=0]{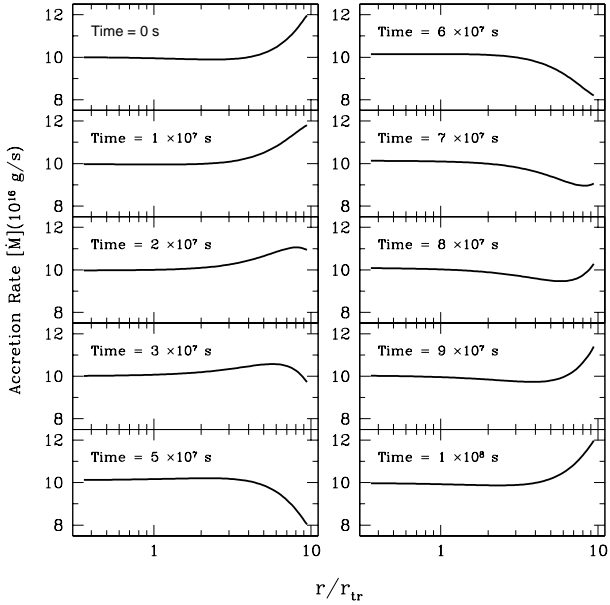}}
\end{center}
\caption{Snapshots of accretion rate versus radii at different times. The perturbation
has been introduced at $10r_{tr}$ and has a time period of 
 $1 \times 10^{8}$ s $ = 2.6 \times (T_{vis})_{tr}$}
\label{movie}
\end{figure}

\section{Time-dependent solutions}

We solve the time-dependent equations for $M = 1.4 M_\odot$, a steady
state accretion rate of $\dot M_s = 1e17$ g/s, $\alpha = 0.1$ and the
efficiency factor $\eta = 0.05$, as we are interested in understanding the
temporal response of the disk to stochastic perturbations initiated at
different radii and for different time periods. Thus at a radius which
we call the perturbation radius $x_p$, we force the accretion rate to
oscillate such that $\dot M = \dot M_s (1 + \Delta \dot M \sin (2 \pi
t/T_p))$, where $T_p$ is the time period of the oscillation and
$\Delta \dot M$ is the normalised amplitude there taken to be a few percent
 for  the simulations.

If the time period $T_p$ is significantly smaller than the 
viscous time-scale at a radius, the perturbation is expected to 
decrease rapidly as it moves inwards. On the other hand, if $T_p$ is
much larger than the viscous one, the perturbation will remain unchanged
at smaller radii. Thus, the region of interest is the behaviour of the
perturbation at radii where the viscous time-scales is of the similar
order as the time period of the oscillations. In this work, to save
numerical time, we introduce the perturbation at a radius where the
viscous time-scale is 3 times the perturbation time period and follow
its evolution to a radius where the viscous time-scale is roughly 
a tenth of the time period. Moreover, we expect interesting physical
behaviour to occur at the transition radius and hence we restrict the analysis 
to perturbations with  time periods
close to the viscous time-scale at that radius which is $\sim 4 \times 10^7$ secs
(Eqn \ref{tvisc}).

To a steady state disk we introduce a perturbation at the prescribed radius
and wait for two times the viscous time-scale to allow for any transient 
behaviour to die down. We then continue the simulation for at least ten times
the time period by which time the system has settled down to unchanging sinusoidal
behaviour. We analyse the results only for the last 8 oscillations. A typical response of
the system is shown in Figure \ref{movie} where time snap shots are shown of
the accretion rate versus radius for one complete oscillation.  The time period
is $ 10^{8}$ sec $ = 2.6 \times (T_{vis})_{tr}$ and is introduced at a radius
which is roughly ten times the transition radius. It is clear from the Figure,
that the perturbation damps out as it moves inwards and amplitude of the oscillation
is rather small in the inner regions. 
\begin{figure}
\begin{center}
{\includegraphics[width=1.0\linewidth,angle=0]{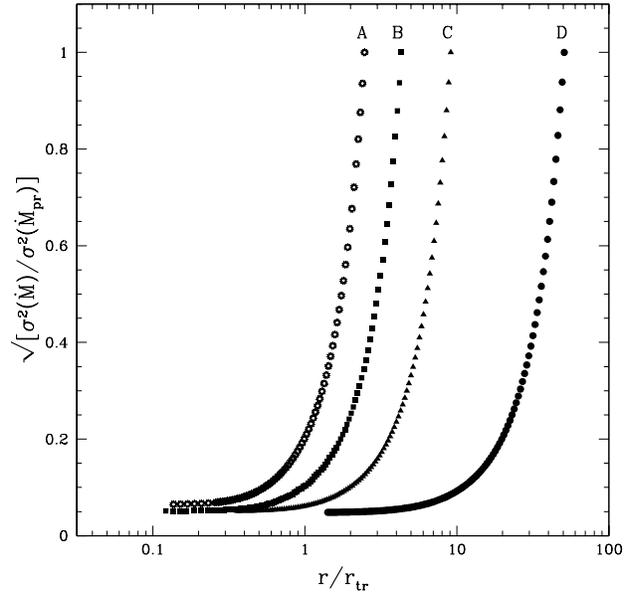}}
\end{center}
\caption{The root mean square (i.e. the square root of the variance $\sqrt{\sigma^2}$) of the
accretion rate perturbation as a function of radius. The r.m.s has been normalised to that 
introduced at the perturbation. The curves are plotted for four different time-periods. For the curves
marked A, B, C and D, the time-periods are $0.8$, $1.3$, $2.6$ and $13.0$ times the viscous time-period
at the Transition radius respectively.The amplitude of the perturbation decreases and then saturates with radius.
}
\label{Time period}
\end{figure}

To quantify the behaviour we compute the
r.m.s $ = \sqrt (\sigma^2 ({\dot M})$ at different radii. In Figure \ref{Time period},
we plot the r.m.s normalised to the r.m.s at the perturbation radius for different time
periods. The behaviour of these curves are expected since for large radii, where the
local viscous time-scale is longer than the time period of the perturbation, the r.m.s
decreases sharply but becomes a constant at radii when the viscous time-scale is shorter.
As mentioned before, we introduce the perturbation at a radius where the local viscous time
scale is 3 times the time period. If the perturbation radius is chosen to be a larger one,
we obtain the same behaviour as shown in the Figure \ref{Time period}, i.e. the curve starts at a larger
radii but has the same shape for the range of radii.
\begin{figure}
\begin{center}
{\includegraphics[width=1.0\linewidth,angle=0]{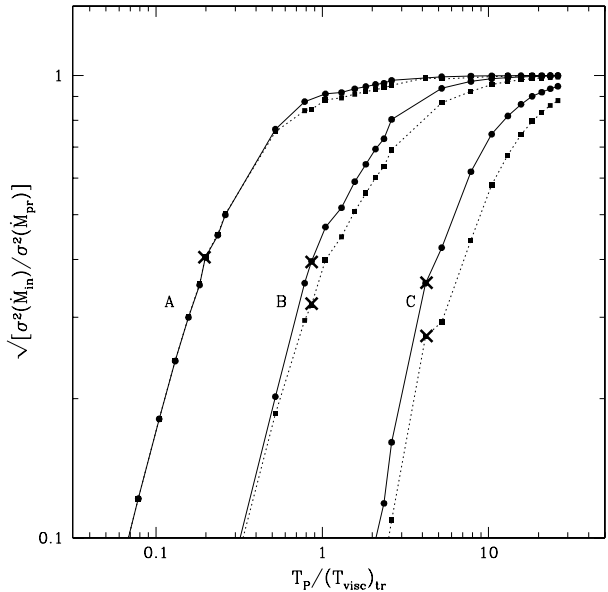}}
\end{center}
\caption{The root mean square variation of the inner accretion rate as a function of
time period. For the curves marked A,B and C, the perturbation radii are $0.3$, $1$ and
$5$ times the transition radius respectively. The dotted lines are for the case when the
X-ray luminosity is artificially kept constant at the steady state value i.e. when there is
no feedback. The crosses on different curves correspond to the viscous time-scale at the corresponding perturbation radius. $\sigma^2 (\dot M_{in})$ is taken to be the variation at a radius where  the 
the viscous time-scale is roughly a tenth of the time period of the
perturbation. When the perturbation is at a radius larger than the transition one, the induced 
inner accretion rate variation is greater than the case without feedback. This indicates that
X-ray irradiation enhances the variability at time-scales larger than $(T_{visc})_{tr}$.}
\label{tp_mdot}
\end{figure}

Another perhaps more illustrative way to show the results is to note the
ratio of the r.m.s of the accretion rate variation in the inner region to
that at the radius of the perturbation i.e. $\sqrt (\sigma^2 ({\dot M}_{in})/\sigma^2({\dot M}_{pr}))$ as 
a function of time period. This essentially tells us how the inner accretion rate (i.e. the X-ray emission) responds
to a perturbation at a given outer radius. The ratio is plotted  for different perturbation radii in Figure 
\ref{tp_mdot}. The ratio is nearly unity when the time period is longer than the
viscous time-scale and is significantly less than one, when the time period is
smaller. To understand the effect of the X-ray irradiation, we also perform a mock simulation
where the X-ray luminosity is artificially kept at the steady state value. The results of such
simulations without X-ray feedback are plotted as dotted lines in Figure \ref{tp_mdot}. When the perturbation
is introduced at a radii less than the transition one (curve marked A) there is as expected no difference
between the cases with and without feedback. However, when the perturbation is introduced at radii larger than
the transition one, the induced variation in $\sigma^2(\dot M_{in})$ is larger for the case with feedback. This
suggests that the effect of X-ray irradiation is to enhance the variability at time-scales larger than the
viscous time-scale of the transition radius.

\section{Discussion}

Our study of a time-dependent  X-ray irradiated disk shows that the response of the inner accretion rate is
different for a perturbation introduced in the X-ray irradiation dominated part of the disk compared to one
at radii less than the transition one. Since the viscous time-scale at the transition radius is $\sim 4 \times 10^7$ secs,
this effect will be observable only by decades long monitoring of these sources. Using the All Sky Monitor (ASM) data of the
RXTE satellite, \cite{Gil05} have reported  breaks in the power spectra of several X-ray binaries indicating that the
power at  frequencies below the break frequency is less than expected. This is in contrast to the results obtained in this work,
which suggests that the variability at low frequency should be enhanced. More seriously, as noted by \cite{Gil05},
the break frequencies occur in the range
$10^{-5}$ - $10^{-7}$ Hz which is much shorter than  the viscous time-scale of the outermost radius and the viscous time-scale
of the transition radius. This indicates that the breaks reported by \cite{Gil05} maybe a different phenomenon unrrelated to the
effects of X-ray irradiation studies here. However given that
the ASM data is not continuous and maybe affected by systematic effects, it is difficult to make a comparison.  
Moreover, the results of this work are based on simplistic assumptions which may need to be
lifted before a meaningful comparison with observations is possible.

We have assumed that the disk is fully ionised but the temperature estimated at the transition radius ($T < 10^4$ K) indicates
that the disk would be partially ionised. The most dominant effect would be that a partially ionized disk  will have  
additional  opacity due to absorption.
The structure of the disk when X-ray irradiation dominates does not depend on the opacity and hence will not be
affected. However, the transition radius may change leading to a different viscous time-scale where the effect of
X-ray irradiation maybe observed. Thus while a time-dependent formalism that computes the ionisation state of the disk
at each radii maybe challenging, it maybe required to predict the quantitative behaviour of the disk. Perhaps a more important effect
which may change the qualitative nature of the result obtained here, is if the irradiated part of the disk is warped, especially if the
warping follows any temporal X-ray modulation. A more sophisticated comprehensive treatment is required to understand
the effects of these aspects.

An important future study would be to predict the time varying  broadband spectrum from such X-ray irradiated disks. This
would require estimating the effective temperature at each radius. While the structure of the disk is affected by the
X-ray irradiation only beyond the transition radius, the effective temperature will depend on the time-dependent X-ray
emission at much smaller radii. The UV/optical emission would have contribution from a span of radii where the effective
temperature will have complex temporal behaviour. The study can predict the relative r.m.s variation as well as time-lags
between the X-ray and different UV/optical bands. For some X-ray binaries there is evidence that the
short time-scale correlation between the X-ray and optical band is complex i.e. although X-ray reprocessing could be a factor,
the optical variability could also be due to other contibutions such as a jet emission \citep[e.g.][]{Mot82,Kan01,Dra14}. This complex short as
well as long term behaviour can be tested using data from the forthcoming X-ray satellite
ASTROSAT \citep{Agr06} which can simultaneously and sensitively measure the optical/UV and X-ray emission from such sources.
Thus, the present work lays the foundation for future studies which will enhance our understanding of the outer
accretion disks of X-ray binaries.  
\section{Acknowledgements}
The authors are grateful to ISRO-RESPOND Program for providing the financial assistance under grant No. ISRO/RES/2/370 for carrying out this work. Three of us (BM, NI and NA) are also highly thankful to IUCAA, Pune for providing the necessary facilties in completing this work.


\label{lastpage}

\end{document}